\documentclass[authoryear,12pt,3p]{jowarticle}

\usepackage[english]{babel}
\usepackage[utf8]{inputenc}
\usepackage{amsmath}
\usepackage{graphicx}
\usepackage[colorinlistoftodos]{todonotes}
\usepackage{natbib}
\usepackage{blindtext}
\usepackage{subcaption}
\usepackage{caption}

\makeatletter
\renewcommand\section{\@startsection{section}{1}{\z@}{-3.25ex plus -1ex minus -.2ex}{1.5ex plus .2ex}{\normalsize\bf}}
\renewcommand\subsection{\@startsection{subsection}{2}{\z@}{-3.25ex plus -1ex minus -.2ex}{1.5ex plus .2ex}{\normalsize\bf}}
\renewcommand\subsubsection{\@startsection{subsubsection}{3}{\z@}{-3.25ex plus -1ex minus -.2ex}{1.5ex plus .2ex}{\normalsize\bf}}
\makeatother

\begin{document}
\begin{frontmatter}
\title{Conformity in Scientific Networks\footnote{This paper was previously distributed and cited with the title ``Do as I Say, Not as I Do, or, Conformity in Scientific Networks''.}}
\author{James Owen Weatherall}\ead{weatherj@uci.edu} 
\author{Cailin O'Connor}\ead{cailino@uci.edu} 
\address{Department of Logic and Philosophy of Science \\ University of California, Irvine}

\date{ }

\begin{abstract}
Scientists are generally subject to social pressures, including pressures to conform with others in their communities, that affect achievement of their epistemic goals.  Here we analyze a network epistemology model in which agents, all else being equal, prefer to take actions that conform with those of their neighbors.  This preference for conformity interacts with the agents' beliefs about which of two (or more) possible actions yields the better result.  We find a range of possible outcomes, including stable polarization in belief and action.  The model results are sensitive to network structure.  In general, though, conformity has a negative effect on a community's ability to reach accurate consensus about the world.
\end{abstract}
\end{frontmatter}

\section{Introduction}

Ignaz Semmelweis, a Hungarian physician, took a post at the first obstetrical clinic of the Vienna General Hospital in 1846. This clinic provided obstetric and early infant care to poor women who were willing to be treated by student doctors.  Its sister institution---the second obstetrical clinic---provided a similar service with midwives in training.  The day of the week determined which of the two clinics patients were referred to.\footnote{For this history see \citet{carter2017childbed,semmelweis1983etiology}.}

But all was not well in the first clinic.  Patients were dying of childbed fever at a rate of about 10\%, while next door the supposedly less expert midwives had a death rate of only 3-4\%.  Worse, women who experienced street births on the way to the clinic were less likely to die than those assisted by student doctors.  In March of 1847, Semmelweis developed a new hypothesis to explain this troubling difference.  When a colleague died of an illness similar to childbed fever after accidentally cutting himself during an autopsy, Semmelweis posited that the student doctors were transferring ``cadaverous particles'' to their patients.  He started requiring students to wash their hands, and the death rate in his clinic plummeted.

Semmelweis published his findings about hand washing, hoping they would change obstetric practices.  In this he was disappointed.  His peers were offended by the suggestion that their hands were unclean, and found his theory of cadaverous particles too far from their theories of disease to credit.  They rejected his new practice despite its clear success when applied. Eventually Semmelweis suffered a nervous breakdown, and died in a mental hospital as a result of a beating he received there.

Long before, in the early 1700s, Lady Mary Wortley Montague, an English aristocrat, traveled to Turkey.\footnote{This history is drawn from \citet{grundy1999lady}.}  While there she encountered the practice of variolation for smallpox, where pus from an infection is introduced to a small scratch.  Those who were variolated usually suffered a very mild form of the disease, and were thereafter immune.  While in Turkey, Lady Mary had her own young son variolated.  

After her return to England, Lady Mary sought to spread the practice, but encountered considerable resistance from English physicians.  They were suspicious of the new treatment, especially because it was being promoted by a woman, who had learned it from women in an infidel country. Even Charles Maitland, Lady Mary's personal physician who had overseen her son's procedure in Turkey, was hesitant to perform variolations under the eyes of fellow British physicians.  Ultimately, she was able to convince Princess Caroline, wife of the future King George II, to have her two daughters variolated in 1722.  After that the practice spread swiftly among English nobility, especially those with personal ties to Lady Mary and Princess Caroline.

That humans tend to conform their actions to those around them has long been documented by psychologists.  \citet{asch1951effects}, in a landmark study, showed that when subjects were placed in a situation where they had to either conform to the obviously incorrect judgments of their peers, or else publicly state a non-conforming, correct judgment, conformity was chosen about 30\% of the time.  Hundreds of subsequent studies have confirmed this tendency towards conformity, though results vary across cultures and from person to person \citep{bond1996culture}.\footnote{In addition, those researching online social networks have found that conformity seems to shape members' choices.  For example, seeing a friend has ``liked'' something on Facebook doubles the chances that a user will ``like" it themselves \citep{egebark2011like}.  
}

In both episodes from the history of science just described, it seems that this tendency to conform influenced the progress of science.\footnote{Of course, we cannot \emph{know} that this was the case, and in the discussion we will consider alternative explanations for the behavior of physicians in these cases.}    Semmelweis's peers did not adopt his new practices, and their choice was bolstered by their conformity with those in their scientific community.  The English physicians who rejected Lady Mary's advocacy for variolation conformed with the actions of practitioners like themselves rather than risk criticism for adopting a new therapy.  Maitland, in particular, was well aware that variolation worked, but still chose to avoid performing it.  Likewise, it was ultimately a preference to conform with the practices of one of the most powerful and influential figures in the country that reversed physicians' resistance to variolation.

Some authors, such as \citet{zollman2010social}, have argued that there are contexts in which conformism tends to \emph{improve} epistemic outcomes, basically because conforming with others can be a heuristic for pooling independent data.  More generally, one can imagine cases in which conformism would be a good heuristic: if you find everyone around you acting in one way, and you are acting differently, it is natural to wonder if they know something you do not.  For this reason, \citet{bikhchandani1998learning} argue that, ``the propensity to imitate is presumably an evolutionary adaptation that has promoted survival over thousands of generations by allowing individuals to take advantage of the hard won information of others'' (152).  But as our case studies indicate, conformity is sometimes bad for scientific belief.  What is going on?

In this paper, we study the interaction between conformism and the epistemic goals of scientific communities by analyzing a network epistemology model.  As we show, the tendency to conform can impact epistemic outcomes in complex ways. But the generally rosy picture on which conformity improves epistemic outcomes is not supported by the results we describe here.  To the contrary, on the whole, conformity tends to produce less successful scientific outcomes via processes similar to those in the Semmelweis and Montague cases.  It also, for some network configurations at least, allows for new phenomena, such as stable polarization in both belief and action.  While there are surely some contexts in which the heuristic identified by \citet{bikhchandani1998learning} and others is useful locally or for individual scientists, its widespread adoption in a scientific community can have detrimental effects.  The character and degree of these effects depend on a number of factors, such as the strength of the preference to conform and the structure of the network in which scientists interact.  

The remainder of the paper will proceed as follows.  In the next section, we will introduce our model and discuss its relationship to other work on conformity in epistemic networks.  The following section will describe the results of our analyses.  We conclude with a discussion of the significance of the results and what sorts of inferences they support.

\section{Network Epistemology and Conformity}

\subsection{The modeling framework}

\citet{venkatesh1998learning} introduce a framework for modeling the social spread of knowledge and belief. \citet{zollman2007communication} imports this framework to philosophy of science to better understand how scientific communities adopt theories, and how their communication structures might influence this process.\footnote{For more work in philosophy of science using this sort of model see \citet{zollman2010epistemic,mayo2011independence,kummerfeld2015conservatism,holman2015problem,holman2017experimentation,Weatherall+etal,OConnor+WeatherallBook,oconnor2017polarization}.  Recently a number of authors have discussed in greater care how these models can and cannot be applied to scientific communities \citep{rosenstock2017epistemic,freyrobustness,freywhatis,borg2017examining}.}

\begin{figure}
\centering
\includegraphics[width=.5\textwidth]{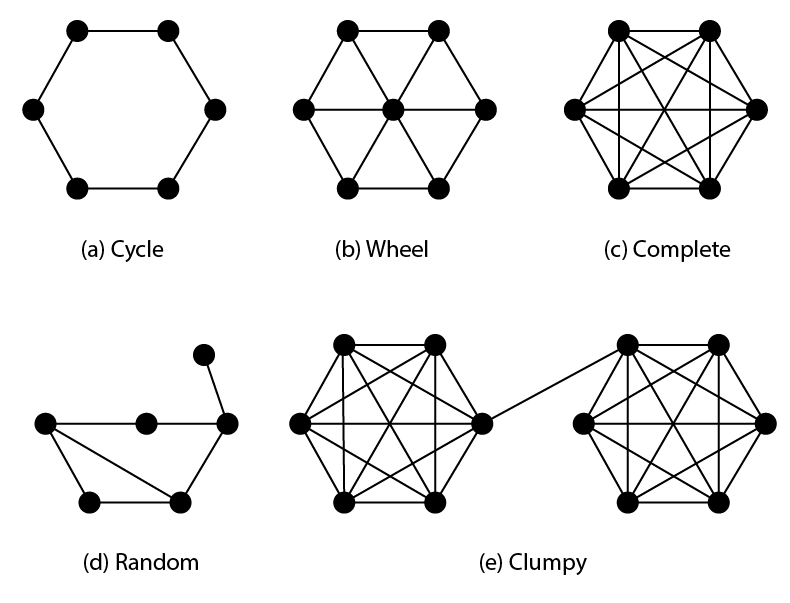}
\caption{Five possible configurations for epistemic networks.}
\label{fig:conformc}
\end{figure}

The framework assumes that there are $N$ agents, or scientists, on a network.  Each agent has a number of symmetric connections to other agents (i.e., if Shareese is connected to Amy, Amy is also connected to Shareese).  These agents choose between actions with different average payoffs.  In doing so they consult evidence that they themselves gather, and evidence gathered by their network neighbors.  Previous authors have considered a number of network structures in such models.  We will look, in particular, at cycle, wheel, complete, and `clumpy' networks, as well as random graphs.  In cycle networks, agents each connect to two others in a ring.  Wheel networks are like these, but with one extra agent who connects to all others.  In complete networks, all agents connect.  In random networks edges are chosen stochastically (more on this later).  Lastly, we consider `clumpy' networks---two complete networks with one link between them---as a structure that will be particularly interesting later on.   These configurations are shown in figure \ref{fig:conformc} for groups of size 6 (and 12 for the clumpy network). 

To represent the epistemic problem agents face, these models employ what is called a `bandit problem'.  The idea is that learners encounter the equivalent of a slot machine (or `bandit') with some number of arms.  Each arm generates a payoff of 1 with a characteristic probability, $p$.  The problem is to choose the  arm with the highest expected payoff.  This type of situation might represent theory choice in a scientific community or a choice between alternative therapies in medicine, for example.  In the Semmelweis case, these choices would be `adopt hand washing' and `do not adopt hand washing', and the success rate would track patient survival.  Following \citet{zollman2010epistemic}, we will initially assume actors face a two-armed bandit problem, where arm A (the All Right arm) pays off with probability $p_A = .5$ and Arm B (the Better arm) pays off with probability $p_B = .5 + \epsilon$.  So arm B is better (wash hands, or variolate), but the actors are uncertain about whether this is the case or not.  (Of course, in the real Semmelweis case, the probabilities of patient survival would be $.9$ for not washing and $.97$ for washing.  The important thing here is that one action is more successful, and gathered evidence reflects this.)   Later, we consider a multi-armed bandit problem, where scientists choose between three or more possible actions.

At the beginning of any simulation of this model, agents start with randomly generated credences about the two arms.  In particular, we assume each agent has two \emph{beta distributions}, one representing their credences about arm A, and the other for arm B.  A beta distribution is a continuous probability distribution whose shape is determined by two parameters, $\alpha$ and $\beta$ (the details of how this works will not be crucial to understanding our results).\footnote{The value of a beta distribution on $x = [0,1]$ is $f(x, \alpha, \beta) = \frac{\Gamma(\alpha+ \beta)}{\Gamma(\alpha)\Gamma(\beta)}x^{\alpha-1}(1-x)^{\beta-1}$ where $\Gamma(y) = (y-1)!$.  An intuitive understanding of the parameters $\alpha$ and $\beta$ is as representing the number of successes and failures garnered in a sample from a bernoulli random variable.  That is, if an agent pulls arm A five times and gets 3 successes, $\alpha = 3$ and $\beta = 2$.  The distribution then specifies probabilities over values of the probability of success for A given this set of results.}  An agent with such a distribution thinks that all probability values, $p$, for an arm are possible, but that some are more likely than others.  We initialize agents beliefs by randomly selecting each of these parameters from the set $[0,4]$.\footnote{While this choice is arbitrary, we follow \citet{zollman2010epistemic}.  These values correspond to an agent who has gathered relatively little evidence, and so will tend to have beliefs that can be updated relatively quickly.}  Each agent then determines which of the two arms they believe has a higher expectation value---the mean expected probability for the arm to pay off---which is equal to $\frac{\alpha}{\alpha+\beta}$.\footnote{Notice that on the intuitive understanding of a beta distribution this is the number of previous successes divided by total number of tests of the arm.}

Every round, an agent pulls their preferred arm some number of times $n$.  This might correspond to a doctor trying a therapy they think more promising (variolation).  Over time, agents update their beliefs about the two arms based on the results of their experiments and those performed by their network neighbors.  We assume they are perfect Bayesians, and use strict conditionalization to update their beliefs.  Since their credences are represented by beta distributions, this is fairly simple.  For an experiment with $n$ draws, and $s$ successes, the new distribution will have parameters $\alpha+s$ and $\beta+n-s$.  Whenever this updating leads an agent to favor a new arm, they switch behavior.

Over time, such a community will tend towards consensus \citep{zollman2010epistemic}.  The actors involved will all come to correctly believe that action B is better, or else to incorrectly believe that action A is better.  The parameter values of the model and network structure will determine how likely these two outcomes are \citep{zollman2007communication,zollman2010epistemic,rosenstock2017epistemic}.

\subsection{A model of conformity}

We described above two episodes from the history of science where conformism seemed to influence the progress of science.   Now, we introduce a model meant to explore this possibility.  This model is very similar to the one just described, but with the extra feature that scientists consider more than just the expected payoff of each action in deciding which to perform.  Instead, they combine their expected payoffs for pulling each arm with further payoffs they expect to get for conforming with those in their network.  

We introduce a conformity parameter $k$, to capture the relative strength of these preferences. What $k$ tracks is how much payoff actors derive from conforming behaviors to their neighbors. Let $a$ be the number of actor $i$'s neighbors who performed action A in the last round, and let $b$ be the number of actor $i$'s neighbors who performed action B in the last round. Then the payoff that agent $i$ expects to receive for performing action A  $(u_i^A)$ is:
\[u_i^A = k(a-b) + \left(\frac{\alpha^A_i}{\alpha^A_i+\beta^A_i}\right),\]
where $\alpha_i^A$ and $\beta_i^A$ are the parameters for agent $i$'s beliefs concerning action $A$.  This says that the agent's expected payoff is a sum of their expectation for arm A, $\left(\frac{\alpha^A_i}{\alpha^A_i+\beta^A_i}\right)$, plus a positive payoff for each agent they conform with ($k*a$) minus a cost for the agents they fail to conform with ($k*-b$).\footnote{As will become clear, we restrict $k$ to be non-negative.  This means our actors cannot have an anti-conformist preference. Observe that as we have set things up, negative payoffs for failing to conform are possible.  If one wished, one could define the same decision problem with only positive payoffs by performing an appropriate affine transformation.}    The expected payoff for arm B is calculated in the same way, except that agents get payoffs for conforming with those who choose B.  Each agent performs the action that they determine to have the higher expected payoff.  Notice that while an agent in this model can come to care arbitrarily little about the truth as $k$ increases, they never discount it entirely. This choice is to maintain the epistemic network character of the models.

There are several choices that we might have made differently.  For instance, we measure payoffs for conformity using the number of neighbors performing each action. As a result, agents with fewer connections tend to be less influenced by conformism, because the total payoff they can receive from conforming (or not) is smaller than for agents with more connections.  We might instead have scaled the conformity payoff using the percentage of neighbors taking each action.  As a robustness check, we ran simulations of this alternative model and found essentially identical results across parameters modulo scaling the value of $k$.  We conclude that this choice has little consequence for the conclusions we wish to draw.

Another choice concerns how agents perform their expected payoff calculation.  Agents do not have beliefs about their neighbors' dispositions to conform. Neither do they attempt to represent their neighbors' epistemic states, or to update their beliefs about those states in light of the evidence that they know their neighbor to have received on previous rounds.  Instead, they simply predict that their neighbors will repeat whatever action they performed in the last round.  Thus, conformity in this model amounts to a kind of imitation of past behavior, and not an attempt to make sophisticated predictions about future behavior.  One could certainly imagine agents with more sophisticated representations of their neighbors, though the resulting model would be much more complicated.  Our choice fits with a picture like that from \citet{bikhchandani1998learning} where humans employ a simple heuristic towards conformity.

\subsection{Related results}

In the next section, we will investigate how this addition to the model influences the emergence of beliefs in a scientific network.  Before that, we will survey previous models of the effects of conformity on the epistemic state of actors on a network, focusing on the philosophy of science literature.\footnote{Economists often investigate conformity using models where agents play a Beauty-contest game.  The goal is to choose an action (perhaps pick a price) with a double goal of matching some state of the world, and also coming as close as possible to the average choice.  There is some similarity to what we do here, as at least some of these models involve networks of agents who gather data from the world, and then make decisions based on both desires for accuracy and conformity \citep{hellwig2009knowing,myatt2011endogenous,colombo2014information}.  They find that the desire for conformity can increase the desire of agents to seek information that is well known.  Our model deviates from these in that we focus on a two-choice problem.  In addition, our agents share data and evidence, as is appropriate for a model of a scientific community.} 

\citet{zollman2010social} argues that conformity may sometimes be a good thing in networks of agents who attempt to choose one of two possible beliefs or theories.  He supposes that each agent gets private information at the beginning of a simulation that with probability $1-\epsilon$ reveals the true state of the world.  Based on this evidence, agents publicly state their opinions about which belief is better.  In subsequent rounds, agents adopt the belief that matches the majority of their network neighbors.  As he shows, this sort of belief conformity often increases the chance that agents end up with correct beliefs, though network structure matters, as does the initial dependability of the agents' information. 

Zollman's result depends on the fact that conformity leads to an aggregation of independent samples of data.  It is related to the Condorcet jury theorem, which established that majority voting based on independent opinions with a greater than 50\% chance of being right will lead to more accurate outcomes the larger the group \citep{condorcet1976essay}.  Our model differs in that we do not assume agents start with good priors.  This, notice, corresponds well to situations where a new, successful scientific belief or practice has been developed (hand washing or variolation) and some individuals are initially skeptical.  We also consider actors who can share data directly rather than spread information via the relatively poor method of publicly stating opinions. This, again, corresponds well to scientific communities.  For these reasons, as we will see, the sorts of benefits to conformity that Zollman identifies will not be as germane, and the detriments will be more apparent.

Previous authors have developed models that suggest a less optimistic picture of conformity.  \citet{mohseni2017} present a model of actors on a network who try to determine which of two states obtains.  These actors gather information from the world in the form of a draw from a distribution that favors the true state.  Like the agents in \citet{zollman2010social}, rather than sharing evidence directly, actors in this network share opinions about which state obtains.  Each round one actor samples the world and then decides on an opinion to state based on both their belief about which state is the right one, and on a desire to conform with the previous statements of neighbors.  Each agent has unknown tendencies towards conformity.  Upon hearing an opinion, the other agents in the network use Bayesian updating to change their beliefs based on the likelihood the actor is conforming with neighbors versus stating their true belief.\footnote{They are able to do this, in part, because each agent `dies' after stating an opinion and is randomly replaced by a new agent.  This creates a `public posterior' that all agents share.}  

These authors prove that in the long run such a network will always arrive at accurate credences about which state of the world obtains. In the short run, however, network structures influence the group's success at determining truth.  Cycle networks, where agents are not highly connected, are particularly conducive to truth since agents are more likely to say what they really think.\footnote{In addition, as we will describe later, cycle networks often have perfectly symmetric social influences, which increase chances that agents make statements based on their beliefs.}  In star networks (where each agent links with one central individual) they find that stated opinions tend to be particularly uninformative, because of the effects of conformity.  As described above, our models make different structural assumptions, and assume a much lower level of rationality for actors.  Even so, some of our results reflect those from \citet{mohseni2017}, adding some robustness to their claims.  We also show, however, that agents can reach stable false beliefs as a direct result of conformism.  This suggests that a central result of their paper, that in the long run networks converge to true beliefs despite conformity, is not robust across modeling choices.

There is another, very influential, literature worth mentioning, which addresses \emph{information cascades}.  \citet{banerjee1992simple} first pointed out that sub-optimal herding behavior, involving significant behavioral conformity, can emerge via a process where actors make rational choices on the basis of group member's behavior.  This happens when a `cascade' of individuals infer that public behavior reveals private information.  They may then decide to conform to this widespread behavior, even if most individuals have information suggesting it is not the right choice.  This literature, though, focuses more on how rational actors may end up conforming as a result of ultimately misleading social information, rather than how innate conformist tendencies impact on epistemic groups.  (See also \citet{bikhchandani1992theory}.)  \citet{berger2018shared} look at human subjects herding behavior, and find that herding is more common when actors share social identity.  These, results, though, may arise from conformity bias, rather than rational updating on the basis of public information.  This is to say that both information cascades and conformity bias can generate patterns of behavior that are very similar, though they arise from different considerations.

\section{Results}\label{sec:}

\subsection{General Analysis}

The addition of conformist tendencies to the network epistemology models discussed above alters the sorts of stable outcomes that are possible.  Recall that without conformity, scientists develop a shared consensus, either correct or incorrect, that completely guides their behavior.  We will call these outcomes either \emph{Correct Consensus} or \emph{Incorrect Consensus}, respectively.  At correct consensus, all agents believe that the success rate of arm A is lower than that of the better arm B, and they all perform action B. At incorrect consensus, all agents believe that the success rate of B is less than that of the worse arm A, and perform action A.  These outcomes are generally stable in the long run.\footnote{Note, however, that stability is never guaranteed in this model, because of its inherent stochasticity.  It is always possible that a run of misleading data will shift agents' beliefs.} 

Both of these sorts of outcomes remain possible when we include conformity, but we also see new possibilities.  First, outcomes are possible that look like consensus in that all scientists take the same action, but where some secretly believe the other action to be the better one.  This can mean that scientists all take the better action, but some erroneously prefer the worse one; or that all scientists take the worse action, but some believe the better one to be, indeed, better.  This latter example is reminiscent of Maitland, Lady Mary's physician: while he knew the practice of variolation worked, he was hesitant to perform it under the eyes of fellow English physicians.  We will call these outcomes \emph{Correct with Disagreement} or \emph{Incorrect with Disagreement}.  Notice that behaviorally, these two outcomes are identical to the first two, but diverge in terms of actors' beliefs.  This is not possible in the model without conformity, since agents always perform the action that they believe has the best payoff.

\begin{figure}
\centering
\includegraphics[width=.4\textwidth]{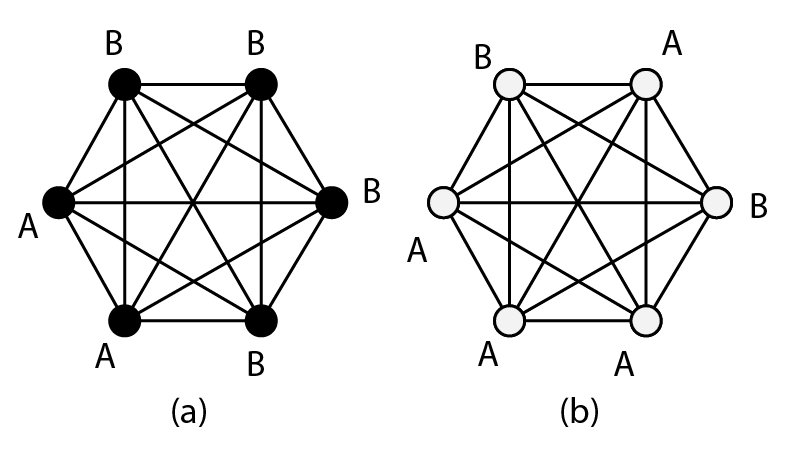}
\caption{With conformity stable outcomes are possible where some individuals do not believe they are taking the better action.  Letters near nodes represent the beliefs the agents prefer.  The color of the node represents their action, black for B and white for A.}
\label{fig:conformdisagree}
\end{figure}

Figure \ref{fig:conformdisagree} shows examples of these outcomes.  These are complete networks with six individuals wherein a) all individuals take action B, as represented by the black nodes; and b) all individuals take action A as represented by the white nodes.  However, in each network some individuals privately hold beliefs that do not match their actions, as represented by the letters next to each node.  Given their conformist tendencies, they will never test the other action.

\begin{figure}
\centering
\includegraphics[width=.4\textwidth]{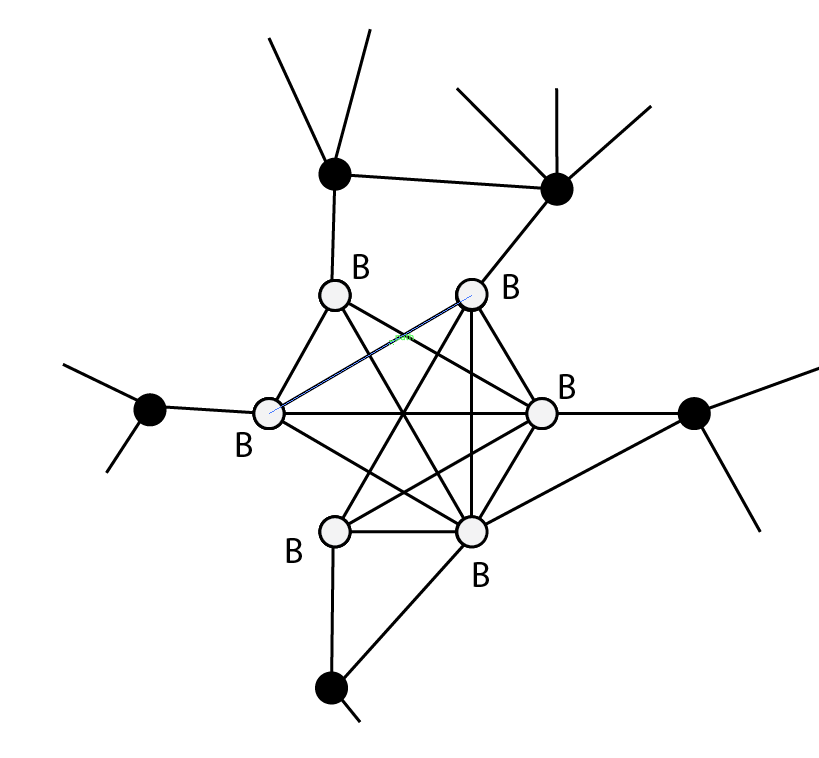}
\caption{With conformity stable outcomes are possible where an entire subgroup holds correct beliefs, but takes the worse action.}
\label{fig:conformodd}
\end{figure}

It is even possible for some subgroups within a community to all take the worse action despite \emph{every} member of that subgroup secretly knowing that B is the better action.  Imagine a subgroup who are tightly connected, and perform action A, where each member is also loosely connected to a larger population performing B.  Every individual will become increasingly certain of the real, underlying success rates of the two arms.  In other words, the means of their beta distributions will approach .5 and $p_B$, and the standard deviations will become smaller and smaller as they continue to observe evidence.  Thus all members of the subgroup will develop correct beliefs, but since they conform with one another, they will not deviate from the group action.  Figure \ref{fig:conformodd} shows an example of what such an outcome might look like.

There is one last stable outcome that emerges as a result of conformism.  We will call this outcome \emph{polarization}, because it involves subgroups within the network where members of one group take action A, and members of the other group take action B.\footnote{The term `polarization' has been used in different ways across the social sciences. We are referencing the general phenomenon where two subgroups hold stably different beliefs, or even diverge in belief, over the course of deliberation.  \citet{oconnor2017polarization} show how this could happen in models like those presented here, i.e., capturing aspects of a scientific community, but where actors do not conform.  Rather, they distrust evidence from those who hold different beliefs from their own.  \citet{bramson2017understanding} overview the literature modeling polarization.}  

Polarization can only emerge for some network structures.  Consider the cycle.  Imagine a temporarily polarized outcome where half the individuals in a cycle choose A and the other half B.  Those on the borders of these groups are under equal pressure to conform either way because they have one neighbor taking each action. Furthermore such an individual gets information about both actions, meaning that over time they develop accurate beliefs.  As this happens they will come to eventually take action B.  This process guarantees that polarized groups are unstable.  In the complete network all agents are under identical conformity pressures, and in a case of transient polarization they get information about both theories, again making polarization unstable.  In the wheel network, things are a bit more complicated, but polarization is never stable there either.

\begin{figure}
\centering
\includegraphics[width=.4\textwidth]{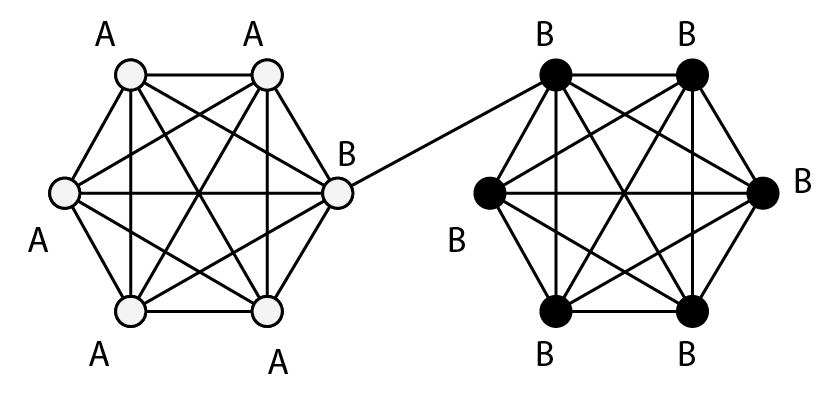}
\caption{With conformity stable polarization outcomes are possible where connected subgroups hold different beliefs.}
\label{fig:conformpolar}
\end{figure}

In some other networks, we do find stable polarization.  As an extreme case, consider the clumpy network.  Here, each subgroup is so tightly connected, and the links between them so minimal, that conformity can stabilize separate actions, even though at least one actor in the `A' group will come to have correct beliefs via her link to the `B' group.  Figure \ref{fig:conformpolar} shows an example of what this might look like, again with node colors representing action B (black) or A (white) and letters representing beliefs.  Notice that the one agent who connects to the B group has correct beliefs, but because she does not switch actions due to conformity she never ends up spreading these correct beliefs through the group.  Random networks can also have the right sorts of cliquish subgroups to get stable polarization.  As we discuss below, these outcomes are found in particular in small world networks, which tend to contain cliques \citep{watts1998collective}.  Many human networks have this small world structure \citep{onnela2007structure,newman2001structure}.  In these cases, the spread of a good idea or practice again can be stymied by social influences.

\subsection{Simulation Results}

To further investigate the effects of conformity on the beliefs and behaviors of our epistemic agents, we run simulations of the model, varying a number of parameters.\footnote{We use simulations here because analytic results would be intractible for this model.  Network models, in general, are difficult to analyze because individuals are in highly asymmetric scenarios.}  We focus on communities of size $N = 10$, except when noted otherwise.  We test different levels of conformity, $k$.\footnote{We studied values of $k$ ranging from $0$ to $20$, but focus on on values $\leq 1$ since increasing $k$ further did not strongly influence results.}  We also vary the probability of payoff for the better arm B, $p_B$, representing a range of scenarios from those where distinguishing the better action is difficult, to those where the two actions have such different success rates that they are easy to tell apart.  ($p_A$, remember is always .5.)\footnote{We focused on simulations where $p_B =.51, .55, .6, .7$.}  In addition, we varied the number of trials scientists ran each time they consulted the world, $n$.\footnote{We considered values of $1, 5, 10, 20, 50, 100$.  This parameter influences how likely it is that a particular set of data will support the better action, B.  When $n$ is small, there are many spurious results, and when $n$ is large there are fewer.  Previous authors have found that this parameter can strongly influence the ability of agents in this sort of model to reach true beliefs \citep{rosenstock2017epistemic}, though this parameter will not be particularly important to the results we state here.}  For each combination of parameter values, we ran 1024 simulations for 10,000 rounds of testing and updating.  This generated outcomes where the community had reached effectively stable beliefs and actions, so that it could be categorized into one of the five outcomes outlined above.\footnote{Generally networks reached stable outcomes in $<<1,000$ rounds, but we ran them for longer to confirm stability. An exception was that in sparse networks, such as the cycle and some small world networks, for low $p_B$ convergence took a long time.  We avoid sharing results where simulations may not have reached stable outcomes.}

\begin{figure}
\centering
\includegraphics[width=.8\textwidth]{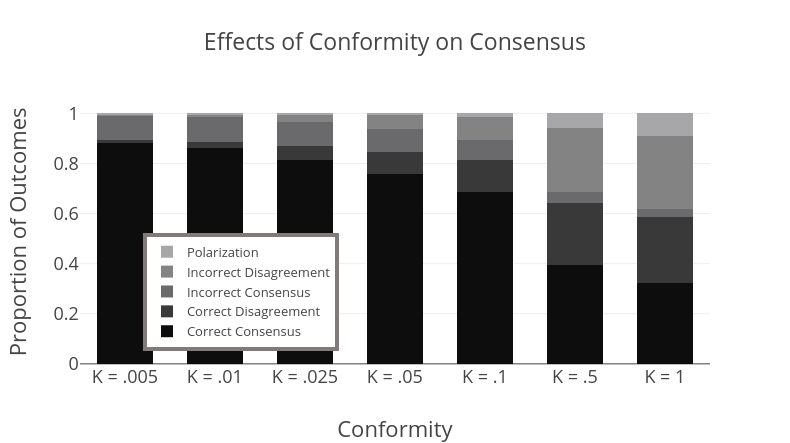}
\caption{Conformity increases outcomes where some actors do not behave according to their beliefs, decreases outcomes where actors take the better action, and increases polarized outcomes. Results are for $G(N,q)$ networks with $N = 10$, $p_B = .55$, $n = 5$, $q = .5$.  The general trend is stable over all tested parameters. }
\label{fig:randomconform}
\end{figure}

We begin by considering how the conformity parameter, $k$, influences outcomes in epistemic networks.  Figure \ref{fig:randomconform} shows outcomes for simulations averaged over random connected networks with ten nodes.\footnote{We ran 10,016 simulations for each combination of parameters of random graphs we considered, to increase our confidence that we were getting a reasonable sample of both possible networks and possible outcomes for each network.}   To construct the networks considered in this figure, we used the \citet{gilbert1959random} $G(N,q)$ random graph algorithm.\footnote{This algorithm is closely related to the Erd\"os-R\'enyi random graph model \citep{erdos1959random} by which one samples from the collection of all graphs with $N$ nodes and $M$ edges, with uniform probability.  It is usually called $G(n,p)$, but we did not wish to confuse readers by using the same variables for different parameters.}  In this algorithm, one begins with $N$ nodes and constructs a graph by considering each possible (undirected) edge linking any two nodes, and including it in the graph with probability $q$.  It is known that for $q=.5$, this algorithm generates all possible graphs on $N$ nodes with equal probability.  Of these, we limit attention to connected networks (i.e., ones where all individuals are connected by some path).\footnote{Since the probability distribution over all networks generated by this algorithm is uniform, restricting to only connected networks still samples from all (connected) networks with equal probability.}  

There are several trends to pick out here.  First, as conformity increases, both disagreement outcomes---where there is behavioral regularity but some individuals hold non-consensus beliefs---increase.  This makes sense.  The more actors care about conformity, the more likely it is that they ignore their private beliefs to do the same thing as their peers.  In contrast, as $k$ increases, both of the consensus outcomes become less likely.  With $k = 0$, all outcomes were one of the two consensus outcomes, with $96\%$ converging to the true consensus.  With $k = .1$, consensus of either sort emerged only about $16\%$ of the time, though in $99\%$ of cases, all agents behaved as if they had reached consensus (i.e., all members took the same action).   Second, conformity makes polarization possible, though polarized outcomes are rare here because only in networks that have the right clique structure can polarization arise.   

Third, the greater the conformity parameter, the less likely, in general, that the community chooses the correct action.  When $k = .0005$, either Correct Consensus or Correct Disagreement is reached 96\% of the time, while incorrect outcomes occur 4\% of the time on average. (Polarization did not occur for small $k$.) When $k = .1$, the correct outcomes occur only 55\% of the time, and incorrect ones 44\%.  The remaining 1\% of the time the groups are polarized, meaning that some portion of the individuals take the worse action.  While these numbers are for particular parameter values, the trend was visible across the data set (modulo a few exceptions discussed later)---more conformity means less successful behavior. 

Why is this the case?  Imagine a group that cares only about conformity, and not about truth at all.\footnote{We never consider this possibility; $k = 20$ is the closest we get.}  This would amount to a pure coordination game among the actors on the network.  Given random starting beliefs, and thus random starting actions, we should expect connected groups to be equally likely to head towards better or worse choices.  There is nothing to break symmetry between these choices.  All actors care about is matching their neighbors, and neither action gives a better payoff beyond this desideratum.  On the other hand, with no conformism, groups of inquirers tend to learn to take the better action (because it is better, and so yields higher payoffs on average).  Increasing conformity siphons away some outcomes where groups would eventually all reach truth, and moves towards the situation where either choice is equally likely.\footnote{Observe that by this analysis, one would expect the probability of arriving at an outcome where all agents perform action B is bounded from below by .5 across networks structures.  This is consistent with our results for all networks we consider in which polarization is not possible.  The possibility of polarization, however, leads to worse outcomes still, because in such cases different subgroups can evolve to different outcomes essentially independently.}

\begin{figure}[htbp]
\centering
\begin{minipage}{0.5\textwidth}
  \centering
\includegraphics[width=0.99\textwidth]{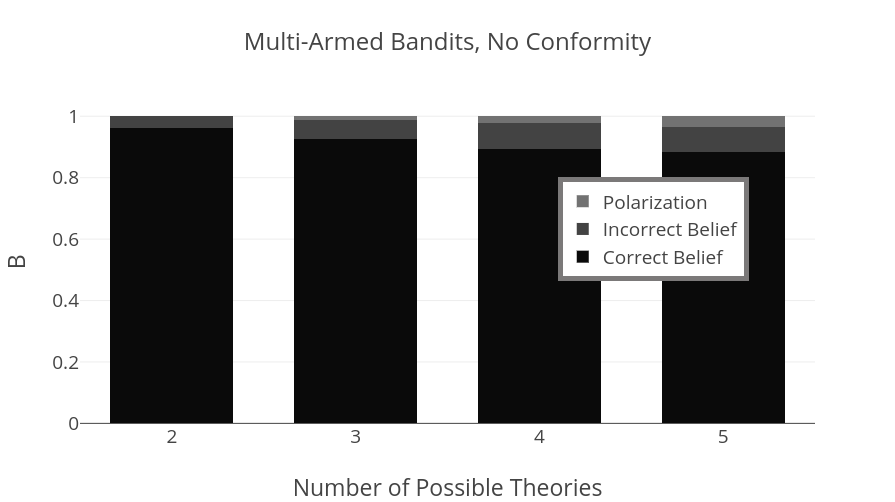}
\subcaption[first caption.]{$k = 0$}\label{fig:1a}
\end{minipage}%
\begin{minipage}{0.5\textwidth}
  \centering
\includegraphics[width=0.99\textwidth]{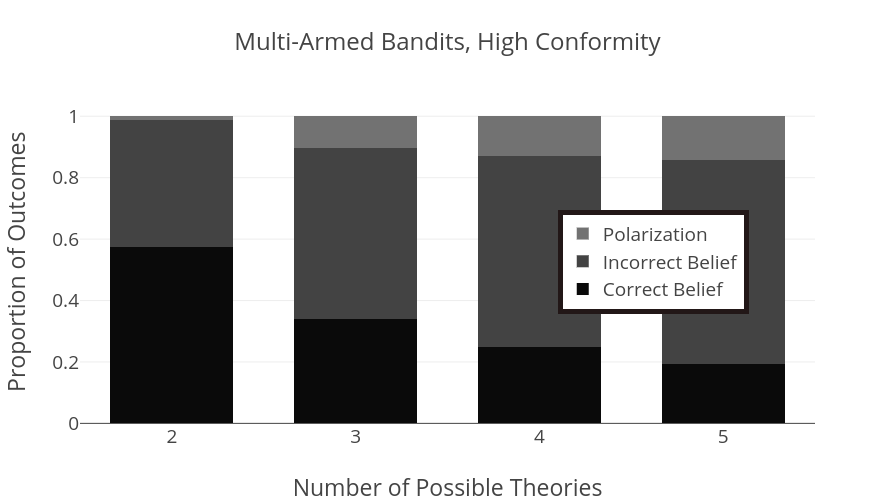}
\subcaption[second caption.]{$k = .5$}\label{fig:1b}
\end{minipage}%

\caption{When scientists test more theories, the effects of conformity are more serious.  Results are for $G(N,q)$ random networks with $N = 10$, $P_B = .55$, $n = .5$, and $q = .5$.} \label{fig:multiarmed}
\end{figure}

This analysis, that varying $k$ is similar to interpolating between the standard Bala-Goyal model and a coordination game on a network, suggests the hypothesis that for large $k$, agents are sometimes getting the right answer for the wrong reason: that is, one of the two actions is chosen essentially at random, without preference for which action yields a higher payoff.  To expand on this hypothesis, we ran simulations for multi-armed bandit problems, where the agents are confronted with more than two possible actions.  In particular, we looked at such models where all arms were successful with probability $p = .5$, except for one better arm, with probability $p_B = .55$.  As expected, for large values of $k$, agents facing multi-armed bandit problems do substantially worse, converging to the true action approximately as often as one would expect if they were choosing one of the n arms with equal odds.  For instance, with $k=.5$, in a three-armed bandit problem agents converge to the better action $34\%$ of the time; in a four-armed bandit problem $25\%$ of the time; and in a five-armed bandit problem $19\%$ of the time.  Contrast this with the case when $k=0$, where adding arms decreases performance, but where, even with five arms, agents converge to the better action more than $88\%$ of the time.  Figure \ref{fig:multiarmed} shows this trend for high and low $k$ as one varies the number of arms.\footnote{Results are for $G(N,q)$ random networks with linking probability $q = .5$. For this figure $n = .5$, population size $10$.  Note that there is `polarization' in the models where $k=0$.  This occurs when actors settle on multiple arms where $p=.5$ and evidence therefore does not discriminate between them.}  We collapse outcomes into correct, incorrect, and polarization. The poor performance of the agents in solving multi-armed bandit problems in the presence of conformity is particularly troubling, given that scientists very often face multiple hypotheses of various effectiveness. 

\begin{figure}
\centering
\includegraphics[width=.8\textwidth]{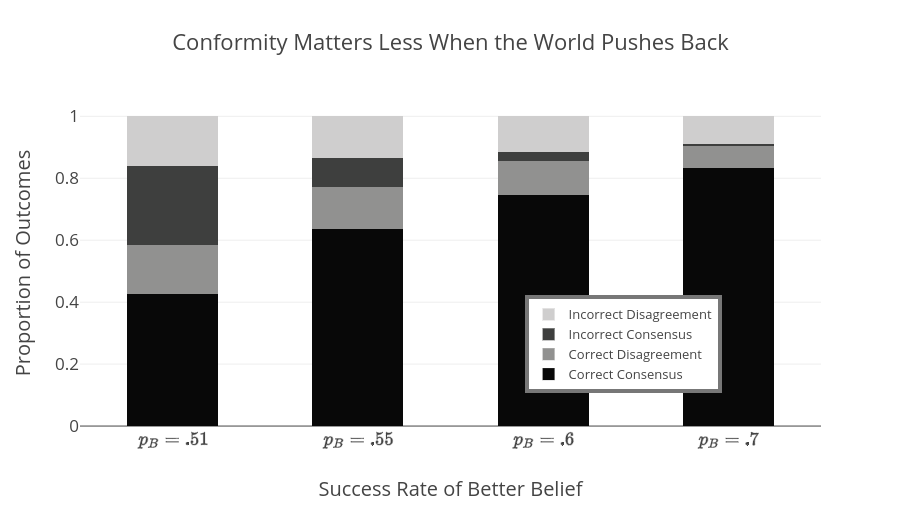}
\caption{When actions matter to payoffs, conformity matters less. Results are over all size 10, complete networks.}
\label{fig:Pbconform}
\end{figure}

In cases where the expected payoff of the better action, $p_B$, is higher, conformity has a less detrimental effect.  Figure \ref{fig:Pbconform} shows this.  These data are the average outcomes across all runs with size 10 complete networks, but the trend is general across the data set.  When $p_B$ increases, making the payoff difference between the choices bigger, the community tends to end up making better choices (and holding more accurate beliefs), despite conformity.  This is in part because when $p_B$ is higher, the data shared between scientists is less likely to spuriously support the worse theory, so there is more data supporting accurate beliefs. But it is also because when payoff differences between the theories are more significant, even actors who care about conformity are more likely to buck consensus and chose the better theory (to get a higher payoff).

If we compare the Semmelweis and Montague case studies, we can see why the latter might be more conducive to the spread of true beliefs along these lines.  While Semmelweis's hand washing practice had very serious implications to real-world payoffs, these implications were removed, to some degree, from the doctors themselves.  They might see patients die, but they were at no risk of death themselves if they failed to wash their hands.  When Princess Caroline variolated her daughters, there was a smallpox outbreak spreading through London.  The possibility of an imminent, grisly death for an individual or their loved ones was likely a contributing factor in overcoming the hesitation some might feel towards a new treatment.  If we consider other, contemporary cases where individuals conform to sets of false beliefs, often (if not always) it is the case that these false beliefs have relatively little influence on these people's lives, at least in the immediate future.  In the case of climate change, for instance, the impact of burning fossil fuels is removed in time, meaning that holding false beliefs is safe (for now).  When it comes to vaccines, herd immunity protects anti-vaxxers from the consequences of their choices.  In the case of evolutionary theory, there are essentially no day-to-day disadvantages that accrue to those who do not believe in it.   In related empirical work, \citet{baron1996forgotten} shows how in simple conformity experiments, like those from \citet{asch1951effects}, actors were less likely to conform when told the task was important and offered money for correct answers.\footnote{For hard tasks, this effect reversed, presumably because uncertain people were depending on their peers to help them reach the right answer.}

We noted above that if we sample uniformly from random connected graphs, polarization appears but it is uncommon.  This is because polarization only appears in networks with cliques, like those in the ``clumpy'' network.  In clumpy networks, polarization appears consistently when $k$ is sufficiently large in 25\% to 50\% of simulations.  This result conforms with the oft-discussed idea that real-world polarization is related to the existence of echo chambers and so-called ``filter bubbles'' \citep{Pariser,flaxman2016filter} on the internet, whereby people are exposed only to others who share their point of view and thereby become entrenched in their opinion.  Polarization in clumpy networks tends to be more likely in cases where the two arms are more difficult to distinguish (i.e., for lower values of $p_B$) and where the quality of the evidence gathered in each round is relatively poor (i.e., the number $n$ of draws per round is low). 

As noted above, small world networks, which resemble many real-world social networks, tend to exhibit cliques, i.e., small subgroups of tightly connected agents who are weakly connected to others in the network \citep{watts1998collective}.  Unsurprisingly, then, we find high levels of polarization in small world networks with conformity.

We construct small-world networks using the Watts-Strogatz algorithm.\footnote{This algorithm begins with a regular ring with $N$ nodes of degree $K$, where $K$ is some positive even number less than $N$ and greater than $\ln(N)$.  To be clear: a cycle as we have discussed it above is a regular ring of degree $2$: each agent is connected to two neighbors.  In a ring of degree $4$, meanwhile, each agent would be connected to their two neighbors as in a cycle, but also to their neighbors' neighbors.  And so on.  Then, the algorithm randomly rewires some connections to create an irregular network. For each node $n_i$ in the network, with $i=1,\ldots,N$, and each edge from $n_i$ to $n_j$, for $i < j$, with probability $b$ delete that edge and create a new edge between $n_i$ and any other node $n_k$, where $n_k$ is drawn with uniform probability from the nodes, aside from $n_i$ itself, to which $n_i$ is not already connected.}  This algorithm randomly generates graphs, with the features of those graphs characterized by three parameters: $N$ (network size), $K$, and $b$.  Increasing $K$ increases the average number of connections between agents.  Increasing $b$ rapidly decreases the average path length.  We consider small world networks with $N=50$, varying the average degree $K$ from $4$ to $16$, and considering $b=.05,.1,.15,.2$.\footnote{We also considered smaller networks, but found that for small $N$, one could not vary $K$ enough to find dependencies while remaining in the ``small world'' regime of small $K$ and low $b$. We considered only $p_B=.55$ and $n=5$, because these simulations took a long time to run.}  Under the effects of conformity, for lower $K$, i.e., on average, fewer connections between agents throughout the network, we find more polarization.  This is because the sparser networks have more clique structures.\footnote{We confirmed that these results were stable by running simulations for a sampling of parameter values for 100,000 rounds and confirming that the proportions of outcomes did not change from the standard 10,000 round cases.}  

\begin{figure}
\centering
\includegraphics[width=.8\textwidth]{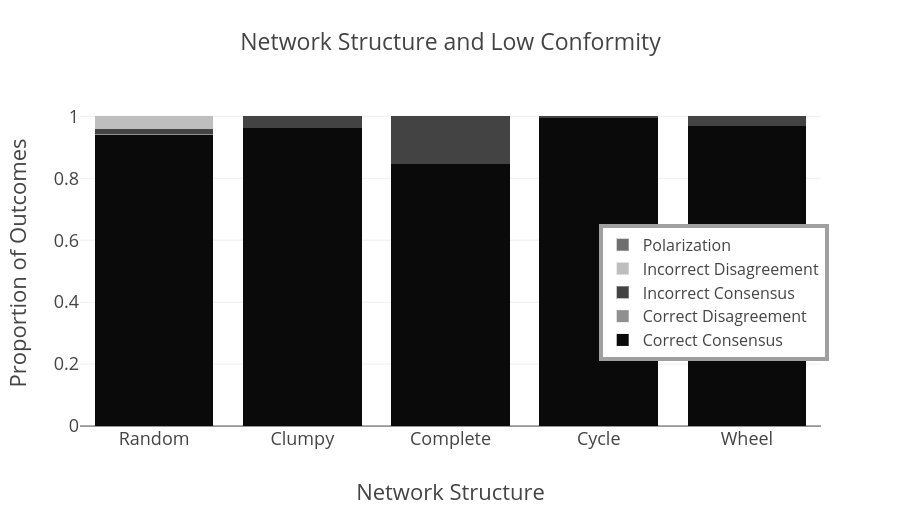}
\caption{Average outcomes for different network structures when conformity is low, $k=.0005$, $p_B=.55$, $n=5$, $N=10$, and $q = .5$.}
\label{fig:CSL}
\end{figure}

More generally, the likelihood of the various possible outcomes tends to vary across network structures.  These effects are complicated because network structure influences the likelihood that an epistemic network arrives at true and false conclusions even without conformity \citep{zollman2007communication,zollman2010epistemic,holman2015problem,kummerfeld2015conservatism,rosenstock2017epistemic}, and that same structure simultaneously affects the influence of conformism.  To tease these effects apart, consider various network structures under a very low conformity value and a very high one.  Figure \ref{fig:CSL} shows outcomes for the different networks when $k = .0005$.\footnote{Random refers to $G(N,q)$ networks with $q = .5$, aka a uniform sampling over all connected networks of size 10.}  As we can see, there are some differences between the networks, but they are relatively minor.  In particular, the complete network is less likely to reach correct consensus than the other three.  This is due to the Zollman effect, where strings of misleading results are sometimes shared to every member of a complete network, leading the entire group to preemptively choose the worse belief \citep{zollman2007communication,zollman2010epistemic}.

\begin{figure}
\centering
\includegraphics[width=.8\textwidth]{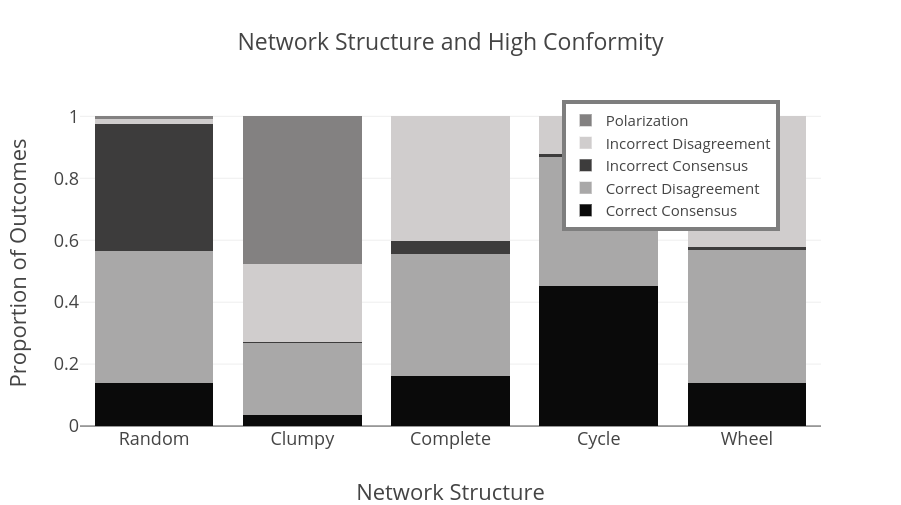}
\caption{Average outcomes for different network structures when conformity is high, $k=.1$, $p_B=.55$, $n=5$, $N=10$, and $q = .5$.}
\label{fig:CSH}
\end{figure}

Compare these results with those in figure \ref{fig:CSH}, which shows values when $k=.1$ (conformity is relatively high), for the same parameter values.  Now, the networks yield very different outcomes.  As already discussed, there is a large amount of polarization in the clumpy networks. Another notable difference is that correct actions are now much more common in the cycle network than the other networks.  This is because at borders where beliefs are spreading, individuals tend to choose beliefs based on the evidence they collect since they are under equal pressure to conform from two neighbors.  In the complete and clumpy networks, on the other hand, there is strong pressure to conform for every agent, leading to worse beliefs.

One might expect the wheel network would be more truth conducive, given its similarity to the cycle.  But for these parameter values the wheel converges to the better act (i.e., True Consensus or Correct Disagreement) only 57\% of the time compared to 87\% of the time for the cycle.  Why might this be?  In the wheel network, the central agent is both the most influential and also under the most social influences, meaning conformity has a strong effect.  This echoes findings from \citet{mohseni2017} that star networks---which are like a wheel, but without the peripheral connections---inhibit successful beliefs.  

This may sound surprising given the case study we have considered in which Princess Caroline, a highly-connected, highly-influential individual, was instrumental in spreading a new, valuable practice.  Princess Caroline was in some ways similar to an agent at the center of a wheel.  But there is an asymmetry here: Princess Caroline was highly influential, but because of her unique social position, she was under less pressure to conform with others.  Our models do not account for the possibility that some individuals might be influential without being highly influenced.  So our models do not capture a relevant aspect of that case that should make a difference.  Note that this case also raises an interesting question that we will defer to future work: under what circumstances does \emph{intervening} on a particular agent's action (or belief) tend to have the greatest significance?

Before moving to the discussion, we will describe one further set of results.  To test the robustness of our findings, we consider models where actors experiment, or err, in the sense that with some small probability on each round they will make a choice based only on beliefs and not conformity.  So, for instance, an actor who believes B is better, but is choosing A for conformity reasons, will sometimes pick B instead.  This difference means that actors with accurate beliefs will eventually, with enough time, transmit these to their neighbors.  It also means that outcomes where actors stick with an action because of conformity are less stable.  Better beliefs are slowly spreading in the network.  It can then randomly happen that some number of clique members try the other action, meaning that in the next round those wishing to conform will switch actions. 

How does this experimentation influence results in the model?  For low levels, results are largely stable.  Outcomes still emerge where actors hold one belief, but take an action that is not prescribed by it in order to conform.  Small world and clumpy networks still develop polarized groups.  As the level of experimentation increases, however, these outcomes become less likely.  For higher levels of experimentation, groups are increasingly likely to develop accurate beliefs, for the reasons noted above---actors continue to learn about both actions, and there are shocks that can unstick the actors from conformist outcomes.\footnote{Of course, these shocks could lead a clique to move from a good action to a worse action.  This is relatively less likely, though, since the actors' good beliefs in these models push them away from bad actions.}

When actors experiment, does conformity still mean worse beliefs?  In large part, the answer to this is yes.  For the vast majority of networks, under most parameter settings, conformity is a bad thing. For the complete network, though, high levels of conformity combined with experimentation can actually improve behavior. This is a bit of a complicated effect.  The experimentation prevents the group from getting stuck conforming to bad behaviors when they have good beliefs.  The conformity seems to improve outcomes because it, paradoxically, increases transient diversity of beliefs.  Without it, actors in the complete network quickly settle into consensus, and sometimes incorrect consensus, because there is so much sharing of evidence.  With it, some actors maintain minority beliefs and continue to experiment with them long enough for these good beliefs to improve the epistemic state of their neighbors.  

\section{Discussion}

As we have seen, in the presence of conformism, new, stable outcomes are possible in epistemic networks.  The actions and beliefs of agents can come apart, meaning agents with correct beliefs may not act in accordance with those beliefs, thus depriving themselves and their neighbors of valuable evidence.  This can lead to a failure of the community to spread accurate beliefs.\footnote{Readers may be interested in a related paper by \citet{imbert2019improving} who look at the dynamics of deliberating groups to consider how to decrease misrepresentation of one's beliefs due to conformist tendencies.}  As a result, it may even be the case that a significant portion of a scientific community (or all of a subgroup) have accurate beliefs about the world, but nonetheless take suboptimal actions.  Of course, the converse is also true: in some cases, agents with \emph{false} beliefs will nonetheless perform the better action.  But this symmetry must be understood in the context of a background \emph{asymmetry} in favor of truth.  In the absence of conformity, across different network structures, Correct Consensus is the most likely outcome.  Conformism, in general, makes actors less likely to take successful actions.

The results discussed here, taken in conjunction with the historical examples presented above, suggest that conformism will have adverse effects on epistemic outcomes across a range of circumstances.  How can this be squared with \citet{zollman2010social}'s arguments that, in some cases, conformism actually improves outcomes?  In fact, even in our models we can see the sort of benefit identified by Zollman in special circumstances.  Consider, for instance, three agents arranged on a line.  Suppose that initially, agent 1 believes action B is better, but agents 2 and 3 believe action A is better. As agent 1 shares results, agent 2 comes to take action B.  Without conformity, agent 3 will generally not begin to perform action B until agent 2 has generated enough results to convince them.  But with conformity, the fact that agent 2 has changed their behavior may lead agent 3 to adopt the better action before they have strong credences in it.  In some cases we see networks that converge more quickly as a result of this sort of process, with relatively little impact on correct belief.\footnote{In particular, we find that in the cycle network, for fixed values of $p_B$, $n$, and $N$, increasing the value of $k$ can decrease the number of rounds needed to reach a state in which the whole network performs the better action by 25\% or more, while having a minimal impact on the fraction of runs in which true action is achieved.  
}  (And, as noted, for select networks with experimentation, conformity can improve outcomes for a different reason, by improving transient diversity of beliefs.)

But, as we have said, conformity generally impairs a community's ability to develop successful beliefs.  Notice that Zollman considers only situations in which agents do not gather or share evidence over time.  Rather, they act based on (sparse) evidence from initial draws. Conformity in his model is thus the only way to pool independent data points.  In our model, on the other hand, the agents are in a better epistemic situation to begin with: they have direct access to one another's evidence, and they can continue gathering evidence over time.  The fact that they can share evidence in this way tends to wash out any benefit of conformism.  The suggestion is that the benefits to conformism that Zollman identifies are real, but arise only in special cases where conformity is necessary to share information.

An important takeaway from the models discussed here is that, in the presence of conformism, network structure becomes important to the success of a scientific community.  Cliquish structures can lead to stable polarization.  Network structure also determines whether individuals face too much social pressure to risk bucking consensus, or whether they feel free to adopt whatever actions their beliefs support.  Previous authors in network epistemology have argued that the structure of scientific communication influences success by influencing patterns of evidence sharing; the suggestion here is that the structure of social pressures in scientific communities does as well.

Along these lines, one aspect of the model we consider that might not be initially obvious is that the network structure is, in fact, tracking two things.  First, it determines who communicates with whom.  Second, it determines who has social influence on whom.  In other words, people are sharing evidence, and thus often, but not always, sharing beliefs with those whom they also try to imitate.  We believe that this is a realistic aspect of many human interactions, and so studying what happens in such a case is important.  That said, we could certainly imagine a situation in which these two aspects of interaction came apart.  Suppose scientists were equally influenced by the evidence of all community members, but only cared about conforming with some small in-group, for example.  A natural extension to this work might involve testing such cases.

One further thing to note is that when it comes to historical cases in which actors stubbornly maintained false beliefs, there are multiple causal pathways that might explain the observed behavior.  As we have pointed out, tendencies to conform can (and likely do) explain many such cases.  In addition, however, it might be the case that individuals had extremely high priors against a certain belief.  Part of what happened in the Semmelweiss case was that his hand washing practice did not accord with current theories of disease.  It might have been the case that all other doctors had such high priors that hand washing could not possibly have the effects Semmelweis claimed, that even after conditioning on his evidence they did not believe his practice was justified.  

Alternatively, \citet{oconnor2017polarization} consider agents in network epistemology models who treat evidence as less certain when it comes from someone with very different beliefs.  As they show, this also can lead to polarization, regardless of network structure, since actors discount the evidence of those on the other side.  In the Semmelweis case, his peers might have had reason to think him a quack because of his odd practices, and so to discount the data he shared, completely separate from any desires to conform with other doctors.  In other words, there are many pathways that might lead to persistent disagreement between scientists, and which might protect false beliefs.\footnote{Furthermore, these pathways suggest nearly completely different interventions.  See \citet{OConnor+WeatherallBook} for a discussion of this point.}

We conclude by observing that the model discussed here, though in some ways a simple extension of a well-studied network epistemology model, leads to a wide range of phenomena, with complex dependencies on both parameters and network structure.  On the one hand, this means that the model can provide new insights and suggest relationships that might not have otherwise been apparent.  But it also means that the results described must be treated with caution.  Indeed, as our analysis shows, many of the results we have described are \emph{not} robust across changes of parameters or network structure---to the contrary, the insights to be gained come from seeing how these phenomena change as we vary parameters.\footnote{Previous authors have delved into the usefulness of network epistemology models, considering robustness of results especially.  See \citet{freyrobustness, freywhatis,rosenstock2017epistemic}.  As they point out, sometimes small changes in these models lead to significantly different results, raising questions for real world applicability.}  So what, then, can we really glean from the model?  

First, it clearly supports the claim we have already emphasized, that conformism generally has a negative effect on epistemic outcomes.  We see these negative effects as $k$ increases across virtually all parameter values and network structures, and we see similarly negative effects in related models from \citet{mohseni2017}.  In other words, this particular result \emph{is} robust.  Second, with respect to other relationships we identify here, the models indicate that these might, in fact, hold in the world.  In other words, we have a `how-possibly' result that conformist tendencies might, for instance, lead to stable polarization.  This connection is not fully established, of course, especially because the models miss many aspects of real-world epistemic networks.  But these models \emph{can} nonetheless play an important role in directing further theoretical and empirical investigations into the effects of conformity on scientific consensus.

\section*{Acknowledgments}
This paper is partially based upon work supported by the National Science Foundation under Grant No. 1535139.  We are grateful to Jeff Barrett, Aydin Mohseni, and Mike Schneider for helpful conversations related to this manuscript.

\bibliographystyle{elsarticle-harv}
\bibliography{Conformbib} 

\begin{thebibliography}{40}
\expandafter\ifx\csname natexlab\endcsname\relax\def\natexlab#1{#1}\fi
\expandafter\ifx\csname url\endcsname\relax
  \def\url#1{\texttt{#1}}\fi
\expandafter\ifx\csname urlprefix\endcsname\relax\def\urlprefix{URL }\fi

\bibitem[{Asch and Guetzkow(1951)}]{asch1951effects}
Asch, S.~E., Guetzkow, H., 1951. Effects of group pressure upon the
  modification and distortion of judgments. Groups, leadership, and men,
  222--236.

\bibitem[{Bala and Goyal(1998)}]{venkatesh1998learning}
Bala, V., Goyal, S., 1998. Learning from neighbors. Review of Economic Studies
  65~(3), 595--621.

\bibitem[{Banerjee(1992)}]{banerjee1992simple}
Banerjee, A.~V., 1992. A simple model of herd behavior. The quarterly journal
  of economics 107~(3), 797--817.

\bibitem[{Baron et~al.(1996)Baron, Vandello, and Brunsman}]{baron1996forgotten}
Baron, R.~S., Vandello, J.~A., Brunsman, B., 1996. The forgotten variable in
  conformity research: Impact of task importance on social influence. Journal
  of Personality and Social Psychology 71~(5), 915.

\bibitem[{Berger et~al.(2018)Berger, Feldhaus, and
  Ockenfels}]{berger2018shared}
Berger, S., Feldhaus, C., Ockenfels, A., 2018. A shared identity promotes
  herding in an information cascade game. Journal of the Economic Science
  Association 4~(1), 63--72.

\bibitem[{Bikhchandani et~al.(1992)Bikhchandani, Hirshleifer, and
  Welch}]{bikhchandani1992theory}
Bikhchandani, S., Hirshleifer, D., Welch, I., 1992. A theory of fads, fashion,
  custom, and cultural change as informational cascades. Journal of political
  Economy 100~(5), 992--1026.

\bibitem[{Bikhchandani et~al.(1998)Bikhchandani, Hirshleifer, and
  Welch}]{bikhchandani1998learning}
Bikhchandani, S., Hirshleifer, D., Welch, I., 1998. Learning from the behavior
  of others: Conformity, fads, and informational cascades. The Journal of
  Economic Perspectives 12~(3), 151--170.

\bibitem[{Bond and Smith(1996)}]{bond1996culture}
Bond, R., Smith, P.~B., 1996. Culture and conformity: A meta-analysis of
  studies using asch's (1952b, 1956) line judgment task. Psychological bulletin
  119~(1), 111.

\bibitem[{Borg et~al.(2017)Borg, Frey, {\v{S}}e{\v{s}}elja, and
  Stra{\ss}er}]{borg2017examining}
Borg, A., Frey, D., {\v{S}}e{\v{s}}elja, D., Stra{\ss}er, C., 2017. Examining
  network effects in an argumentative agent-based model of scientific inquiry.
  In: International Workshop on Logic, Rationality and Interaction. Springer,
  pp. 391--406.

\bibitem[{Bramson et~al.(2017)Bramson, Grim, Singer, Berger, Sack, Fisher,
  Flocken, and Holman}]{bramson2017understanding}
Bramson, A., Grim, P., Singer, D.~J., Berger, W.~J., Sack, G., Fisher, S.,
  Flocken, C., Holman, B., 2017. Understanding polarization: Meanings,
  measures, and model evaluation. Philosophy of Science 84~(1), 115--159.

\bibitem[{Carter(2017)}]{carter2017childbed}
Carter, K.~C., 2017. Childbed fever: a scientific biography of Ignaz
  Semmelweis. Routledge.

\bibitem[{Colombo et~al.(2014)Colombo, Femminis, and
  Pavan}]{colombo2014information}
Colombo, L., Femminis, G., Pavan, A., 2014. Information acquisition and
  welfare. The Review of Economic Studies 81~(4), 1438--1483.

\bibitem[{Condorcet(1785)}]{condorcet1976essay}
Condorcet, M.~d., 1785. Essai sur l'application de l'analyse à la probabilité
  des décisions rendues à la pluralité des voix.

\bibitem[{Egebark and Ekstr{\"o}m(2011)}]{egebark2011like}
Egebark, J., Ekstr{\"o}m, M., 2011. Like what you like or like what others
  like? conformity and peer effects on facebook.

\bibitem[{Erd\"os and R\'enyi(1959)}]{erdos1959random}
Erd\"os, P., R\'enyi, A., 1959. On random graphs i. Publ. Math. Debrecen 6,
  290--297.

\bibitem[{Flaxman et~al.(2016)Flaxman, Goel, and Rao}]{flaxman2016filter}
Flaxman, S., Goel, S., Rao, J.~M., 2016. Filter bubbles, echo chambers, and
  online news consumption. Public Opinion Quarterly 80~(S1), 298--320.

\bibitem[{Frey and {\v{S}}e{\v{s}}elja(2017{\natexlab{a}})}]{freyrobustness}
Frey, D., {\v{S}}e{\v{s}}elja, D., 2017{\natexlab{a}}. Robustness and
  idealizations in agent-based models of scientific interaction.

\bibitem[{Frey and {\v{S}}e{\v{s}}elja(2017{\natexlab{b}})}]{freywhatis}
Frey, D., {\v{S}}e{\v{s}}elja, D., 2017{\natexlab{b}}. What is the function of
  highly idealized agent-based models of scientific inquiry?

\bibitem[{Gilbert(1959)}]{gilbert1959random}
Gilbert, E.~N., 1959. Random graphs. The Annals of Mathematical Statistics
  30~(4), 1141--1144.

\bibitem[{Grundy(1999)}]{grundy1999lady}
Grundy, I., 1999. Lady Mary Wortley Montagu. Clarendon Press, Oxford.

\bibitem[{Hellwig and Veldkamp(2009)}]{hellwig2009knowing}
Hellwig, C., Veldkamp, L., 2009. Knowing what others know: Coordination motives
  in information acquisition. The Review of Economic Studies 76~(1), 223--251.

\bibitem[{Holman and Bruner(2017)}]{holman2017experimentation}
Holman, B., Bruner, J., 2017. Experimentation by industrial selection.
  Philosophy of Science 84~(5), 1008--1019.

\bibitem[{Holman and Bruner(2015)}]{holman2015problem}
Holman, B., Bruner, J.~P., 2015. The problem of intransigently biased agents.
  Philosophy of Science 82~(5), 956--968.

\bibitem[{Imbert et~al.(2019)Imbert, Boyer-Kassem, Chevrier, and
  Bourjot}]{imbert2019improving}
Imbert, C., Boyer-Kassem, T., Chevrier, V., Bourjot, C., 2019. Improving
  deliberations by reducing misrepresentation effects. Episteme, 1--17.

\bibitem[{Kummerfeld and Zollman(2015)}]{kummerfeld2015conservatism}
Kummerfeld, E., Zollman, K.~J., 2015. Conservatism and the scientific state of
  nature. The British Journal for the Philosophy of Science 67~(4), 1057--1076.

\bibitem[{Mayo-Wilson et~al.(2011)Mayo-Wilson, Zollman, and
  Danks}]{mayo2011independence}
Mayo-Wilson, C., Zollman, K.~J., Danks, D., 2011. The independence thesis: When
  individual and social epistemology diverge. Philosophy of Science 78~(4),
  653--677.

\bibitem[{Mohseni and Williams(2017)}]{mohseni2017}
Mohseni, A., Williams, C.~R., 2017. Truth and conformity on networks~(Working
  Paper).

\bibitem[{Myatt and Wallace(2011)}]{myatt2011endogenous}
Myatt, D.~P., Wallace, C., 2011. Endogenous information acquisition in
  coordination games. The Review of Economic Studies 79~(1), 340--374.

\bibitem[{Newman(2001)}]{newman2001structure}
Newman, M.~E., 2001. The structure of scientific collaboration networks.
  Proceedings of the National Academy of Sciences 98~(2), 404--409.

\bibitem[{O'Connor and Weatherall(2017)}]{oconnor2017polarization}
O'Connor, C., Weatherall, J.~O., 2017. Scientific polarization,
  arXiv:1712.04561 [cs.SI].

\bibitem[{O'Connor and Weatherall(2019)}]{OConnor+WeatherallBook}
O'Connor, C., Weatherall, J.~O., 2019. The Misinformation Age: How False
  Beliefs Spread. Yale University Press.

\bibitem[{Onnela et~al.(2007)Onnela, Saram{\"a}ki, Hyv{\"o}nen, Szab{\'o},
  Lazer, Kaski, Kert{\'e}sz, and Barab{\'a}si}]{onnela2007structure}
Onnela, J.-P., Saram{\"a}ki, J., Hyv{\"o}nen, J., Szab{\'o}, G., Lazer, D.,
  Kaski, K., Kert{\'e}sz, J., Barab{\'a}si, A.-L., 2007. Structure and tie
  strengths in mobile communication networks. Proceedings of the national
  academy of sciences 104~(18), 7332--7336.

\bibitem[{Pariser(2011)}]{Pariser}
Pariser, E., 2011. The filter bubble: How the new personalized web is changing
  what we read and how we think. Penguin, New York.

\bibitem[{Rosenstock et~al.(2017)Rosenstock, Bruner, and
  O'Connor}]{rosenstock2017epistemic}
Rosenstock, S., Bruner, J., O'Connor, C., 2017. In epistemic networks, is less
  really more? Philosophy of Science 84~(2), 234--252.

\bibitem[{Semmelweis(1983)}]{semmelweis1983etiology}
Semmelweis, I.~F., 1983. The etiology, concept, and prophylaxis of childbed
  fever. No.~2. Univ of Wisconsin Press.

\bibitem[{Watts and Strogatz(1998)}]{watts1998collective}
Watts, D.~J., Strogatz, S.~H., 1998. Collective dynamics of
  ‘small-world’networks. nature 393~(6684), 440--442.

\bibitem[{Weatherall et~al.(2017)Weatherall, O'Connor, and
  Bruner}]{Weatherall+etal}
Weatherall, J.~O., O'Connor, C., Bruner, J., 2017. How to beat science and
  influence people, arXiv:1801.01239 [cs.SI].

\bibitem[{Zollman(2007)}]{zollman2007communication}
Zollman, K.~J., 2007. The communication structure of epistemic communities.
  Philosophy of science 74~(5), 574--587.

\bibitem[{Zollman(2010{\natexlab{a}})}]{zollman2010epistemic}
Zollman, K.~J., 2010{\natexlab{a}}. The epistemic benefit of transient
  diversity. Erkenntnis 72~(1), 17.

\bibitem[{Zollman(2010{\natexlab{b}})}]{zollman2010social}
Zollman, K. J.~S., 2010{\natexlab{b}}. Social structure and the effects of
  conformity. Synthese 172~(3), 317--340.

\end{thebibliography}

\end{document}